\documentclass{article}
\usepackage{graphicx}
\usepackage{amsmath}

\begin{document}

\begin{center}

{\Large \textbf{Density of Inertial Particles: Exactly Solvable 2D Models}}

\bigskip
\bigskip

{\textbf{Leonid I. Piterbarg}}

\bigskip
\bigskip

{Department of Mathematics, University of Southern California}

{Kaprielian Hall, Room 108, 3620 Vermont Avenue, Los Angeles, CA 90089-2532}

\bigskip
\end{center}

{\large \textbf{Abstract} }

\bigskip

Inertial particles in 2D  driven by a Gaussian white noise forcing  are considered.
For two examples of the forcing (compressible and incompressible) upper and lower bounds are found for the mean number of caustics as a function of Stokes number.
Efficiency of the bounds is verified by numerical methods.

\bigskip


\bigskip

{\large \textbf{Introduction}}

\bigskip
In compressible fluid flows the density of Lagrangian particles finite at an initial moment can go to infinity in finite time as a flow is evolving. Such density explosions are related to occurence of caustics in the underlying velocity field. One of the most important such phenomena is motion of inertial particles in  turbulence ( a comprehensive review can be found in [1]).

In this work we address the mean frequency (intensity) of caustic occurence $\nu$ in the framework of a stochastic flow modeling Lagrangian motion in a turbulent velocity field. The stochastic flow concept allows one for describing joint statistics of several particles or even a particle continuum at any moment.

The main goal is to introduce two particular models in two dimensions, very simple but not trivial, for which the dependence of $\nu$ on Stokes number $s$ (the ratio of Lagrangian and Eulerian time scales, [2]) can be analytically investigated to a certain extent. Namely, first, lower and upper bounds for $\nu(s)$ are provided yielding an exact asymptotic as $s\to 0$ and, second, $\nu (s)$ is accurately evaluated for a wide range of $s$ by solving a $1D$ parabolic equation numerically.

A general version of a stochastic flow addressed here has been introduced in [3] and assumes finite particle velocities and a white noise type of accelerations. Thus, it is not a Kraicnan model of turbulence widely used in physics of fluids [2]. More exactly, assume that the position $\mathbf r(t, \mathbf a)$ of a particle at time $t$ with the initial position $\mathbf a$ and its velocity $\mathbf v(t,\mathbf a)$ are described by  the following Ito equations
$$
d\mathbf r =\mathbf v dt,\ \ d\mathbf v =-(\mathbf v/\tau) dt +d\mathbf w (t,\mathbf r),\ \ \ \mathbf r(0, \mathbf a)=\mathbf a,\ \ \mathbf v(0, \mathbf a)=\mathbf u_0(\mathbf a) \eqno (1)
$$
where  $\mathbf u_0(\mathbf x)$ is an initial (Eulerian) velocity field, $\mathbf w(t, \mathbf r$) is a random forcing with zero mean which is a Brownian motion in time and a statistically homogeneous random field in  space , i.e.
$$
E\mathbf w(t_1,\mathbf r_1)d\mathbf w(t_2,\mathbf r_2)^T=\min (t_1,t_2)\mathbf B(\mathbf r_1- \mathbf r_2)
$$
where $\mathbf B(\mathbf r)$ is its space covariance matrix, all the vectors are two-dimensional, and finally $\tau$ is the Lagrangian correlation time scale.

In particular, the motion of a single particle in the framework (1) is covered by a Langevin equation for the velocity
$$
d\mathbf v =-(\mathbf v/\tau) dt +\mathbf B(\mathbf 0)^{1/2}d\mathbf w (t)
$$
where $\mathbf w (t)$ is a standard Brownian motion in 2D, thereby the Lagrangian velocity of the particle is simply a well known (two-dimensional) Ornstein-Uhlenback process and its position is an integrated Ornstein-Uhlenback process.

In general the motion of any number $n$ of the particles is described by a diffusion process in $R^{2n}$ for the vector of positions/velocities $(\mathbf r_1, \mathbf v_1, ...,\mathbf r_n, \mathbf v_n )$  with a drift and diffusivity matrix expressed in terms of $\tau$ and $\mathbf B$, [3].

In this work we mostly address the Jacobian
$$
J(t,\mathbf a)=\displaystyle\det\left(\frac{\partial \mathbf r}{\partial \mathbf a}\right) \eqno (2)
$$
and partially  the top Lyapunov exponent (LE)
$$
\lambda = \displaystyle \lim_{t\to \infty}\frac 1t\lim_{|\mathbf a|\to 0}\ln\left(\frac{|\mathbf r(t,\mathbf a)-\mathbf r(t,\mathbf 0)|}{|\mathbf a|}\right) \eqno (3)
$$
which depends on the direction of $\mathbf a$ for anisotropic flows.

 A consideration of  two-particle motion suffices for studying both characteristics .

An important physical meaning of the Jacobian can be seen from the following. Let the initial position $\mathbf a$ be a random variable independent of the flow with pdf $\pi_0(\mathbf a)$, then for the conditional pdf $\pi(t,\mathbf r)$ conditioned on the flow is given by
$$
\pi(t,\mathbf r(t, \mathbf a))=\pi_0(\mathbf r(t,\mathbf a))/J(t, \mathbf a)
$$
Hence, if for some $t_0$ the Jacobian takes zero value, the density becomes infinite at the point $\mathbf r(t_0, \mathbf a)$ what we will call an occurrence of a caustic at this point.

We show that for a certain class of forcings caustics arise with probability one, they form a stationary process in time and sensible bounds for the mean number of caustics are given.

In Eulerian terms equations  (1) cover Lagrangian motion in the Eulerian velocity field $\mathbf u(t, \mathbf x)$ satisfying the following equation
$$
\frac{\partial \mathbf u}{\partial t}+\mathbf u \cdot \nabla \mathbf u +\mathbf u/\tau=\frac{\partial \mathbf w}{\partial t}  ,\ \ \ \ \mathbf u|_{t=0}=\mathbf u_0(\mathbf x) \eqno (4)
$$
By the method of characteristics one can find that for some short interval $(0,t)$ the solution of this equation exists and unique since at the initial moment $J(0, \mathbf a)=1$. However after some time the uniqueness is lost (the Jacobian takes a zero value) due to a very weak dissipation modeled by the last term on the left hand side (LHS). Thus, estimates of the mean of the first moment when $J(t,\mathbf a)$ hits zero would provide us with an idea when (4) looses  uniqueness. In terms of the non-linear wave  theory, one can treat that as the first moment of  wave breaking.

Some of our results can be predicted by simply procceding to dimensionless variables in (4). Let $U$ and $L$ be some typical velocity and length scales respectively.
Changing  $\mathbf u, \mathbf w, \mathbf x, t$ to $\mathbf u/U, \mathbf w/U, \mathbf x/L, t/\tau$ one gets from (4)
$$
\frac{\partial \mathbf u}{\partial t}+s\mathbf u \cdot \nabla \mathbf u +\mathbf u=\frac{\partial \mathbf w}{\partial t}   
$$
where the dimensionless parameter
$$
s=\displaystyle \frac{U\tau}{L}   \eqno (5)
$$
is called Stokes number.

Thus, if $s \to 0$ then the underlying Eulerian velocity field satisfies a linear equation and hence the intensity of caustics should tend to zero. In the opposite case $s\to\infty$ the non-linear term dominates and one may expect that the number of caustics per time  grows indefinitely.

The paper is organized as follows. A similar one-dimensional  problem is exactly solved in Section 1. The solution is essentially used in investigating  2D phenomena. A deterministic case ($\mathbf w =\mathbf 0$) is briefly discussed in Section 2. In Section 3 we derive a system of equations containing $J$ in the general case of an arbitrary homogeneous forcing. For particular forms of the forcing rigorous estimates for the mean number of caustics $\nu (s)$ are found in Section 4. Also $\nu(s)$ is computed by solving a simple $1D$ parabolic equation.  The case of an isotropic forcing (unsolved yet) is mentioned in Section 5.  Conclusions are gathered in Section 6 and some details are brought to Appendix.

\bigskip

{\large \textbf{1. One-dimensional case}}

\bigskip

A part of results from this section is well known from physical literature, [1].
We  formulate them in rigorous form and give more details.

Consider
$$
dx=udt,\ \ du=-(u/\tau)dt+dw(t,x),\ \ \ x(0)=a,\ \ u(0)=u_0(a)     \eqno (6)
$$
with $Ew(t,x)=0$ and
$$
Ew(t_1,x_1)dw(t_2,x_2)=min(t_1, t_2))b(x_1-x_2)
$$
For
 $$
J(t,a)=\frac{\partial x}{\partial a}
$$
one can get by direct differentiation (6) in $a$
$$
dJ=J_1dt,\ \ dJ_1=-(J_1/\tau)dt+Jd\tilde w(t)
$$
where
$$
J_1(t,a)=\frac{\partial u}{\partial a},\ \ \ \ E\tilde w(t)=0,\ \ E\tilde w(t)^2=-b''(0)t
$$
By setting $p=\tau J_1/J$, introducing dimensionless time $t=t/\tau$, and using Ito formula obtain
$$
dp=(-p-p^2)dt+\sqrt{2}sdw \eqno (7)
$$
where $w$ is a standard Wiener process and
$$
s =\sqrt {-b''(0)\tau^3} \eqno (8)
$$.
Worth noting that if the velocity and length scales of the flow are chosen as $U=\sqrt{b(0)\tau}$ and $L=\sqrt{-b(0)/b''(0)}$ respectively, then (8) coincides with the earlier introduced Stokes number (5).

Assume that zeros of $J(t)$ are prime that is typical for stationary processes, then they can be identified with moments of explosion of $p(t)$. 

\textbf{Proposition 1}.

\textit {Process} $p$ \textit{is explosive(e.g. [4]), more exactly}
\[
\quad \lim_{t\rightarrow S-0}p(t)=-\infty ,
\]
\textit{where the explosion time} $S$ \textit{is finite}
\[
\quad P(S<\infty )=1
\]
\textit {and its expectation can be explicitly computed}
$$
E(S)= 2\pi^2 \sqrt{z}M(z)^2,\ \ z=s^{-4/3}/4   \eqno (9)
$$
\textit{where}
\[
 M(z)=\sqrt{Ai(z)^{2}+Bi(z)^{2}},\quad
\]
$Ai(z),Bi(z)$ \textit{are the Airy functions and} $M(z)$ \textit{their
modulus}.

\smallskip
Explosiveness of $p(t)$ follows from the Feller's criteria, [4].

To prove (9) introduce $\mu(x)$ as the mean time to explosion under condition $p(0)=x$, then solve
\[
\quad L\mu(x)=-1,\quad \mu(-\infty )=0
\]
and set
\[
E(S)= \mu(\infty ).
\]
where
$$
 L=(-x -x^{2})\frac{\partial }{\partial x}+s^2 \frac{\partial ^{2}}{\partial x^2} 
$$
 is the generator of (7). Details of computations can be found in [5].

For the purpose of studying $2D$ models the knowledge of the mean explosion time is not enough. Introduce
$$
\bar F(t) = P(S>t)
$$
and let $u(t,x)$ be the same probability under condition $p(0)=x$, then
$$
\bar F(t) = u(t,\infty)
$$
In [6] it is shown that $u(t,x)$ satisfies the initial value problem
$$
\displaystyle  \frac{\partial  u(t,x)}{\partial t}=L_xu(t,x), \ \ u(0,x)=1 \eqno (10)
$$
To ensure uniqueness of solution of (10) we add the natural boundary conditions assuming that the limit $u(t,\infty)$ exists  similarly to $m(\infty)$.
$$ 
u(t,-\infty)=0, \ \ u_x(t, \infty) = 0 
$$
A simplest Euler scheme was used for solving (10) numerically. Some details and validation of the choice of the scheme parameters are given in Appendix. 
Notice that our goal is not to evaluate and minimize the error of the numerical computations, but rather to illustrate that (10) can be solved quite accurately and efficiently with very simple tools.

Graphs of $\bar F(t) $ for few values of $s$ are shown in Fig.1

\begin{figure}[hbp]
\begin{center}
\includegraphics[width=80mm, height=60mm]{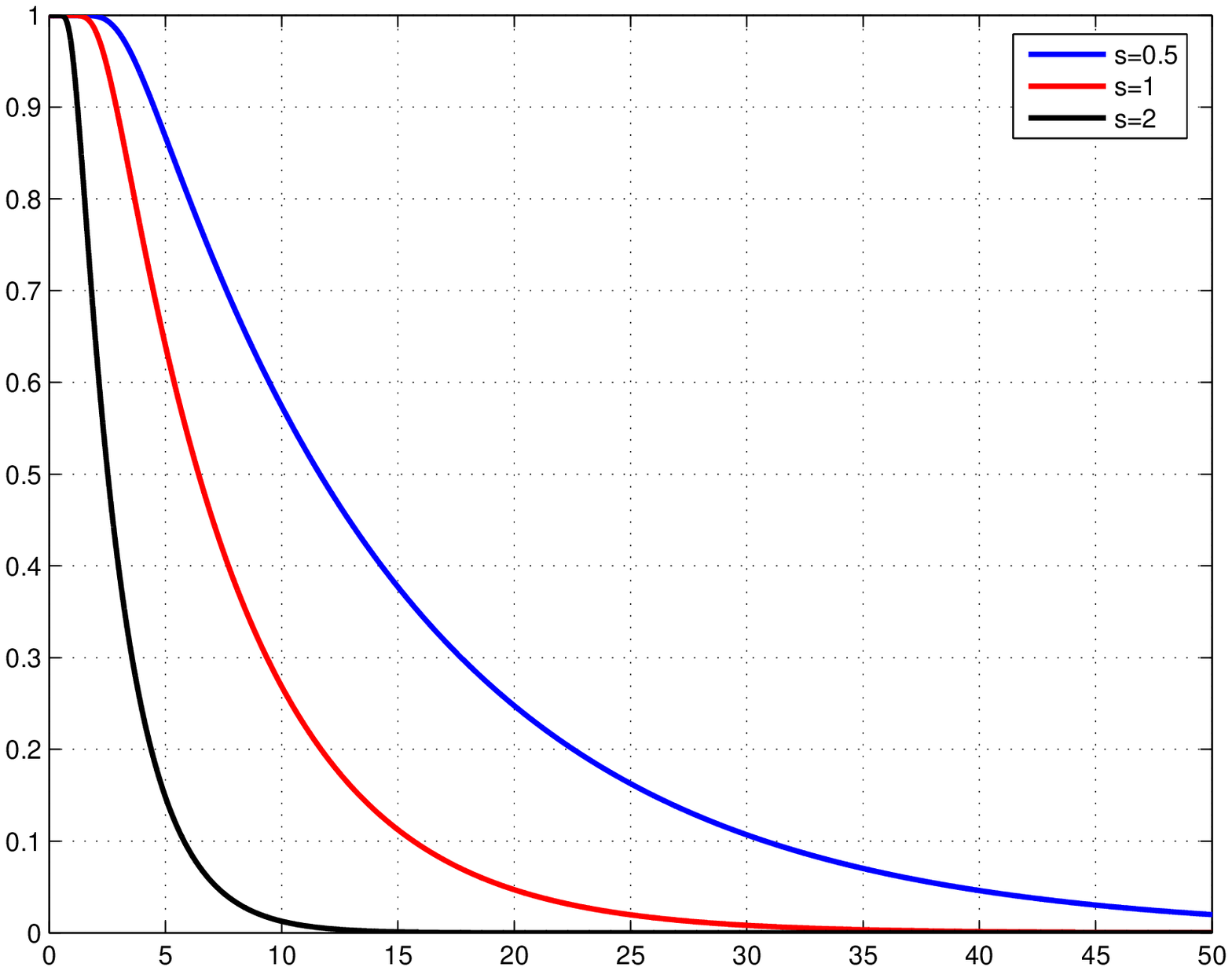}
\end{center}
Figure 1. $P(S>t)$ versus $t$ for $s= 0.5, 1, 2$
 \end{figure}

After explosion process $p(t)$ can be continued by starting over, i.e. by solving same eq. (7) with the initial condition
\[
 \lim_{t\rightarrow
S+0}p(t)=\infty
\]
The positive sign follows from relation $J_1=dJ/dt$ and the fact that zeros of $J(t)$ are prime. Thus $p(t)$ takes different signs on different sides of a zero of $J(t)$.
\smallskip

LE for flow (7)   is expressible in terms of the ergodic mean of $p$
$$
\lambda = \bar p/\tau,\ \ \ \bar p=\displaystyle \lim_{t\to\infty}\frac 1t \int_0^t p(s)ds  
$$
as follows from (3) and definition of $p(t)$, see [3] for details. It also can be exactly found

\textbf{Proposition 2}.

\textit{The ergodic mean of $p$ is given by}
$$
\bar p =  -1+\frac{M^{\prime }(z)}{2\sqrt{z}M(z)}
\eqno (11)
$$

The statement is proven in Appendix and here we just make some comments.
Notice that the density of the stationary distribution $\pi(p)$  of $p(t)$ does exist and is given by
$$
\pi (p)=Ce^{-p^2/2s^2-p^3/3s^2}\int_{-\infty}^p e^{y^2/2s^2+y^3/3s^2}dy  \eqno (12)
$$
where $C$ is a normalized constant.

One can find the following asymptotic
\[
\pi (p)\sim |p|^{-2},\quad p\rightarrow \infty .
\]
Thus, the integral for the invariant mean
$$
Ep=\int_{-\infty}^{\infty}p\pi(p)dp 
$$
is formally divergent, however  if the integral is meant as a Cauchy principal value then it coincides with the RHS of (11).

For bounded functions of $p$ the ergodicity holds true with a conventional interpretation of the integral over the invariant density.

\textbf{Proposition 3}

\textit{Process} $p(t)$ \textit{is ergodic , i.e. for any bounded function} $%
f(p)$
$$
\lim_{T\rightarrow \infty }\frac{1}{T}\int_{0}^{T}f(p(t))dt=\int_{-\infty
}^{\infty }f(p)\pi(p)dp
$$

\bigskip

{\large \textbf{2. Deterministic  2D Case }}

\bigskip
Let us assume $\mathbf w = 0$ in (1) and proceed to dimensionless variables as described in Introduction before eq. (5). As a result the expression for the Jacobian (2) is found in the explicit form
$$
J(t)=Ps^2\left(1-e^{-t}\right)^2 +Qs\left(1-e^{-t}\right)+1
$$
where $s$ is Stokes number defined in (5)
$$
\displaystyle P=\frac {\partial (u_0,v_0)}{\partial (a,b)},\ \ Q= \frac {\partial u_0}{\partial a}+\frac {\partial v_0}{\partial b}
$$
and $(a,b)$, / / ($u_0, v_0)$ are the initial position and velocity respectively related by $u_0=u_0(a,b),\ \ v_0=v_0(a,b)$.

Let
$$
G=\{(P,Q)|\ Q^2>4P\}\bigcap\{(P,Q)|\ P<0\ \rm{or}\ Q<0\}
$$
and $S$ be the first time when $J(S)=0$. Easy to find that
$$
\displaystyle S=\left\{
 \begin{array}{lll}
\displaystyle -\ln \left(1-\frac{s_0}{s}\right) & {\rm if} \ \ (P,Q)\in G\ \ {\rm and}\ \ s>s_0\\ \\
\infty & \  {\rm otherwise}
\end{array}
\right.
$$
where
$$
\displaystyle s_0=\frac{-Q-\sqrt{Q^2-4P}}{2P}
$$
In particular if $Q=0$ and $P<0$ do not depend on $(a,b)$ at all, then $G=R^2$ and $S$ , defined for all $s>0$, does not depend on the initial point as well. Thus, the equation 
$$
\frac{\partial \mathbf u}{\partial t}+s\mathbf u \cdot \nabla \mathbf u +\mathbf u=\mathbf 0 ,\ \ \  \mathbf u(0, \mathbf x)=\mathbf u_0(\mathbf x) \eqno (13)
$$
looses uniqueness exactly at moment $t=S$.
If one defines $u(t,x)$ after $t>S$  as a solution of (13) with the initial condition 
$$
\mathbf u(S+, \mathbf x)=\mathbf u_0(\mathbf x) 
$$
then  $\nu=1/S$ could be interpreted as the intensity of caustics in time.  In Figure 2 we show few curves $\nu(s)$ for different $s_0$ in order to compare them visually with a stochastic case addressed in the next sections.

\begin{figure}[hbtp]
\begin{center}
\hspace{0.8cm}\\
\includegraphics[width=100mm, height=100mm]{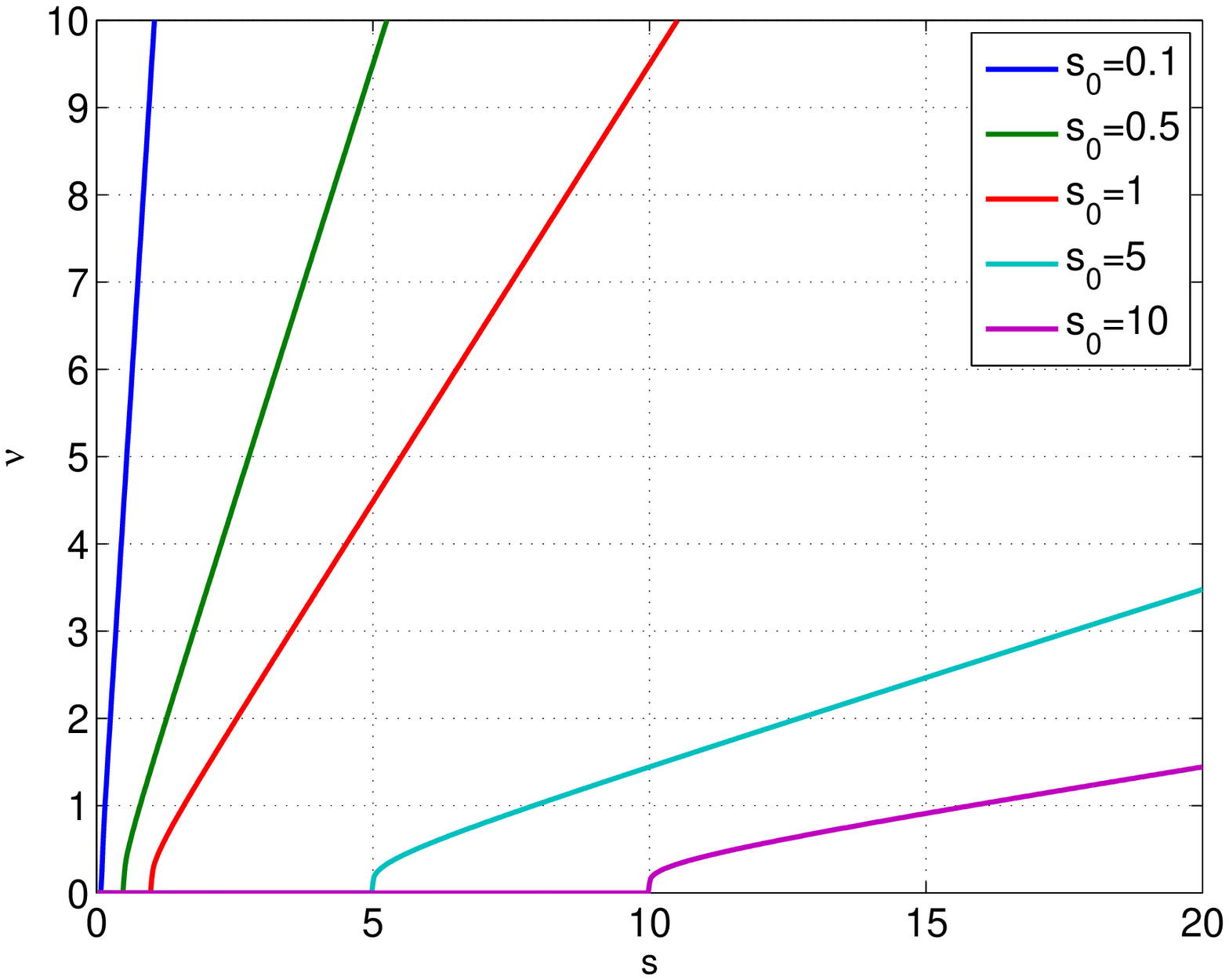}\\
\end{center}
Figure 2. Dependence of the caustic rate $\nu(s)$ on Stokes number $s$ in absence of noise.
\end{figure}
If $S$ depends on the initial point, then  its interpretation in terms of Eulerian set up (13) is not that simple and will not be discussed here.
\bigskip

{\large \textbf{3.  System of Equations for Jacobian in general 2D Case}}

\bigskip

Now we return to the stochastic model (1). In few works it was pointed out that there is a closed equation for the matrix
$$
\mathbf Z =\frac{\partial \mathbf v}{\partial \mathbf a}\left(\frac{\partial \mathbf r}{\partial \mathbf a}\right)^{-1}
$$
Namely, e.g. [3]
$$
\displaystyle d\mathbf Z=-\left(\mathbf Z/\tau+\mathbf Z^2\right)dt +d\left(\frac{\partial \mathbf w}{\partial \mathbf a}\right)
$$
An analysis of that equation led to some general important conclusions, but it is of little help for our purposes because in general it cannot be reduced to efficiently handled scalar equations.

Let us first rewrite  (1) in the coordinate wise form with $\mathbf r =(x,y),  \mathbf a =(a,b)$
$$
\begin{array} {lll}
du=-(u/\tau )dt+dw^{(1)}(t,x,y),\quad dv=-(v/\tau
)dt+dw^{(2)}(t,x,y),\quad dx=udt,\quad dy=vdt \\ \\
x(0)=a,\ \ y(0)=b,\ \ \ \ u(0)=u_0(a,b),\ \ v(0)=v_0(a,b)
\end{array}   \eqno (14)
$$
where
\begin{equation*}
Edw^{(i)}(t,0,0)dw^{(j)}(t,x,y)= b^{ij}(x,y)dt
\end{equation*}
and $b^{ij}$ are entries of $\mathbf B$
The goal is to investigate time behavior of the Jacobian
\begin{equation*}
J(t)\equiv \frac{\partial x}{\partial a}\frac{\partial y}{\partial b}-%
\frac{\partial x}{\partial b}\frac{\partial y}{\partial a}
\end{equation*}
It is not possible to obtain a closed equation for $J(t)$, but it can be included in a system of four equation as it is shown in Appendix.
Namely, for dimensionless time $t=t/\tau$ denoted by the same letter, Jacobian can be represented as
\begin{equation*}
J(t)=\exp \left(\int_{0}^{t}m_{1}(t)dt\right),
\end{equation*}
where dimensionless random function $m_{1}(t)$ is included into the following system
\begin{equation*}
\begin{array}{lll}
dm_{1} & = & (-m_{1}+(m_{4}^{2}-m_{1}^{2}-m_{2}^{2}-m_{3}^{2})/2)dt+sd\eta_1, \\
dm_{2} & = & (-m_{1}m_{2}-m_{2})dt+sd\eta_2, \\
dm_{3} & = & (-m_{1}m_{3}-m_{3})dt+sd\eta_3, \\
dm_{4} & = & (-m_{1}m_{4}-m_{4})dt+sd\eta_4
\end{array}
\eqno(15)
\end{equation*}
$s$ is Stokes number defined similarly to (8) and an exact expression for it is given in Appendix.

To define processes $\eta_i(t)$ we nondimensionalize $w^{(i)}(t)$ and set
$$
\eta_1=w^{(2)}_y-w_x^{(1)},\ \ \eta_2=w^{(2)}_y+w_x^{(1)},\ \ \eta_3=w^{(2)}_x-w_y^1,\ \ \eta_4=w^{(2)}_x+w_y^{(1)},\ \eqno (16)
$$
where  the subs mean derivatives. Thus $\eta$'s are dependent Wiener (non-standard) processes with
a covariance matrix $Ed\eta^{(i)}(t,x,y)d\eta^{(j)}(t,x,y)= \sigma^{ij}_\eta dt$ given by
\begin{equation*}
(\sigma^{ij}_\eta)=\left(
\begin{tabular}{llll}
$-b^{11}_{xx}-b^{22}_{yy}+2b^{12}_{xy}$ & $-b^{22}_{yy}+b^{11}_{xx} $& $b^{12}_{xx}+b^{12}_{yy}$ & $b^{12}_{xx}-b^{12}_{yy}$ \\ \\
$-b^{22}_{yy}+b^{11}_{xx}$ & $-b^{11}_{xx}-b^{22}_{yy}-2b^{12}_{xy} $& $-b^{12}_{xx}+b^{12}_{yy}$ & $-b^{12}_{xx}-b^{12}_{yy}$ \\ \\
$b^{12}_{xx}+b^{12}_{yy}$ & $-b^{12}_{xx}+b^{12}_{yy} $& $-b^{22}_{xx}-b^{11}_{yy}+2b^{12}_{xy}$ & $-b^{22}_{xx}+b^{11}_{yy}$ \\ \\
$b^{12}_{xx}-b^{12}_{yy}$ & $-b^{12}_{xx}-b^{12}_{yy} $& $-b^{22}_{xx}+b^{11}_{yy}$ & $-b^{22}_{xx}-b^{11}_{yy}-2b^{12}_{xy}$ \\ \\
\end{tabular}
\right)\eqno (17)
\end{equation*}
where $b^{ij}$ is a dimensionless version of $b^{ij}$ and  partial derivatives are taken at $x=0, y=0$.

\bigskip

{\large \textbf{4. Two special cases}}

\bigskip

Under certain conditions imposed on the forcing in (14) the first moment $S$ of $J(t)$ to hit zero turns out to be the minimum from  the first explosion moments for two independent 1D processes described by (7).

In this section we assume a zero initial velocity field
$$
\mathbf u_0(\mathbf x) \equiv \mathbf 0 
$$
that implies  zero initial conditions  $m_k (0)=0,\ \  k>1$ in (15)
\bigskip

{\bf Model 1}

\smallskip

Assume
$$
w^{(1)}(t,x,y)= w^{(1)}(t,x),\ \ w^{(2)}(t,x,y)= w^{(2)}(t,y) \eqno (18)
$$
where $w^{(1)}(t,x),  w^{(2)}(t,y)$ are independent and
$$
E\left(w^{(1)}_x\right)^2=E\left(w^{(2)}_y\right)^2 \eqno (19)
$$
From (16,18) it follows that
$$
\eta_3=0,\ \ \eta_4=0
$$
Then from (16,17,19) $E\eta_1^2=E\eta_2^2,\ \ \ E\eta_1\eta_2=0$ and hence 
$$
w_1=(\eta_1+\eta_2)/2,\ \  w_2=(\eta_1-\eta_2)/2
$$
are independent identically distributed Wiener processes.

Next due to the zero initial conditions  $m_3=m_4\equiv 0$. Introduce
$$
p=\frac{m_1+m_2}{2},\ \ q=\frac{m_1-m_2}{2}
$$
By adding and subtracting first two equations in (15) we get two separated equations
$$
dp=(-p-p^2)dt+s dw_1,\ \ dq=(-q-q^2)dt+s dw_2 \eqno (20)
$$
for independent identically distributed processes $p(t)$ and $q(t)$

\bigskip

{\bf Model 2}

\smallskip

In this model we assume
$$
w^{(1)}(t,x,y)= U(t,x+y)+V(t,x-y),\ \ w^{(2)}(t,x,y)= -U(t,x+y)+V(t,x-y) \eqno (21)
$$
with independent $U$ and $V$ such that
 $$
E\left(U'\right)^2=E\left(V'\right)^2 \eqno (22)
$$
From (16, 17, 21, 22) it follows  that
$$
\eta_2=0,\ \ \eta_4=0
$$
and
$$
w_1=(\eta_1+\eta_3)/2,\ \  w_2=(\eta_1-\eta_3)/2
$$
are independent identically distributed Wiener processes since $\eta_1=-2(U'+V')$ and $\eta_3=-2(U'-V')$

Thus $m_2=m_4 \equiv 0$ and introducing
$$
p=\frac{m_1+m_3}{2},\ \ q=\frac{m_1-m_3}{2}
$$
arrive at the same equations (20).

Below we show simulated velocity fields at a particular time moment for both models (Fig.3) where periodic in space forcings were used. 

\begin{figure}[hbtp]
\begin{center}
 \includegraphics[width=70mm, height=70mm]{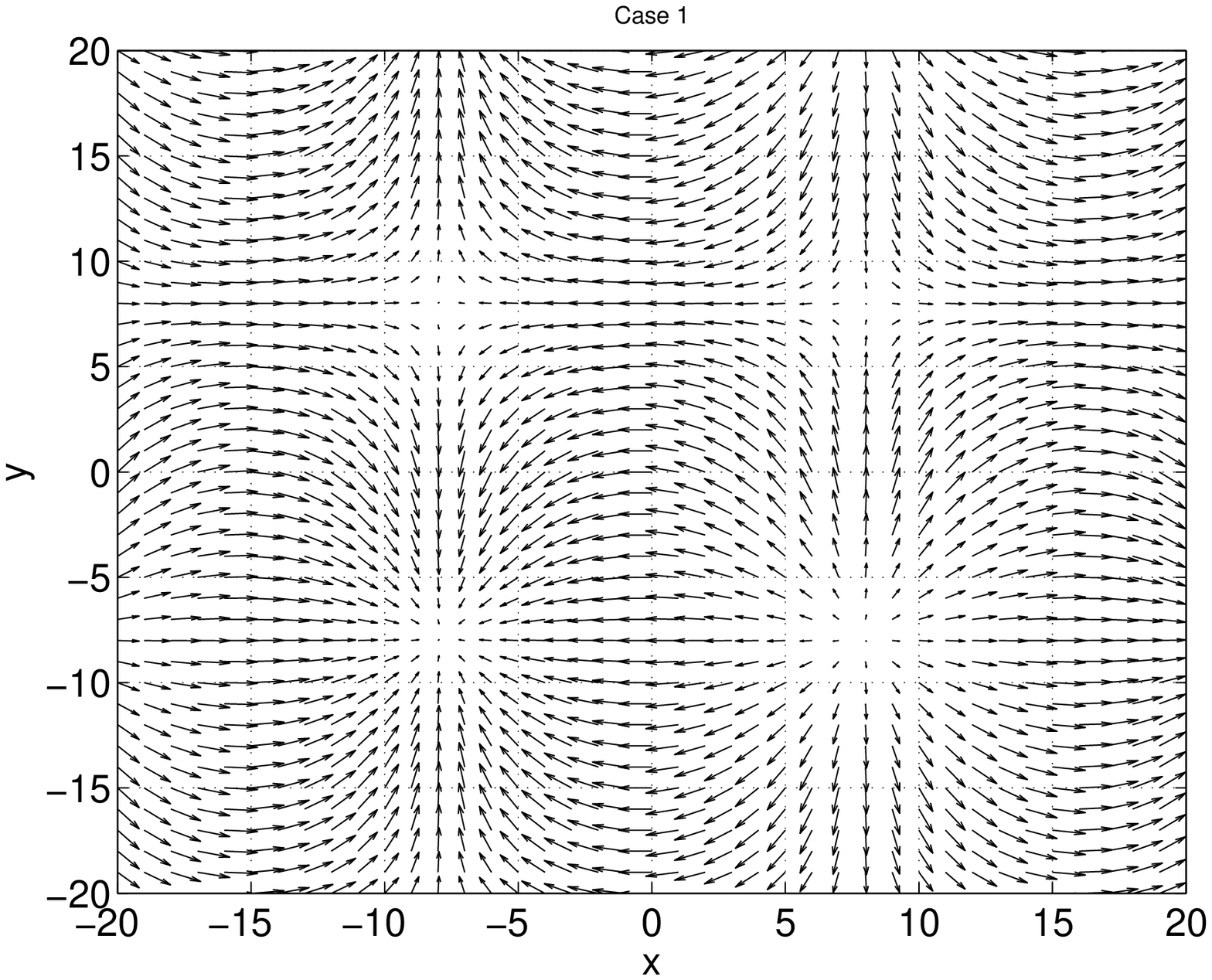}\ \ \ \
\includegraphics[width=70mm, height=70mm]{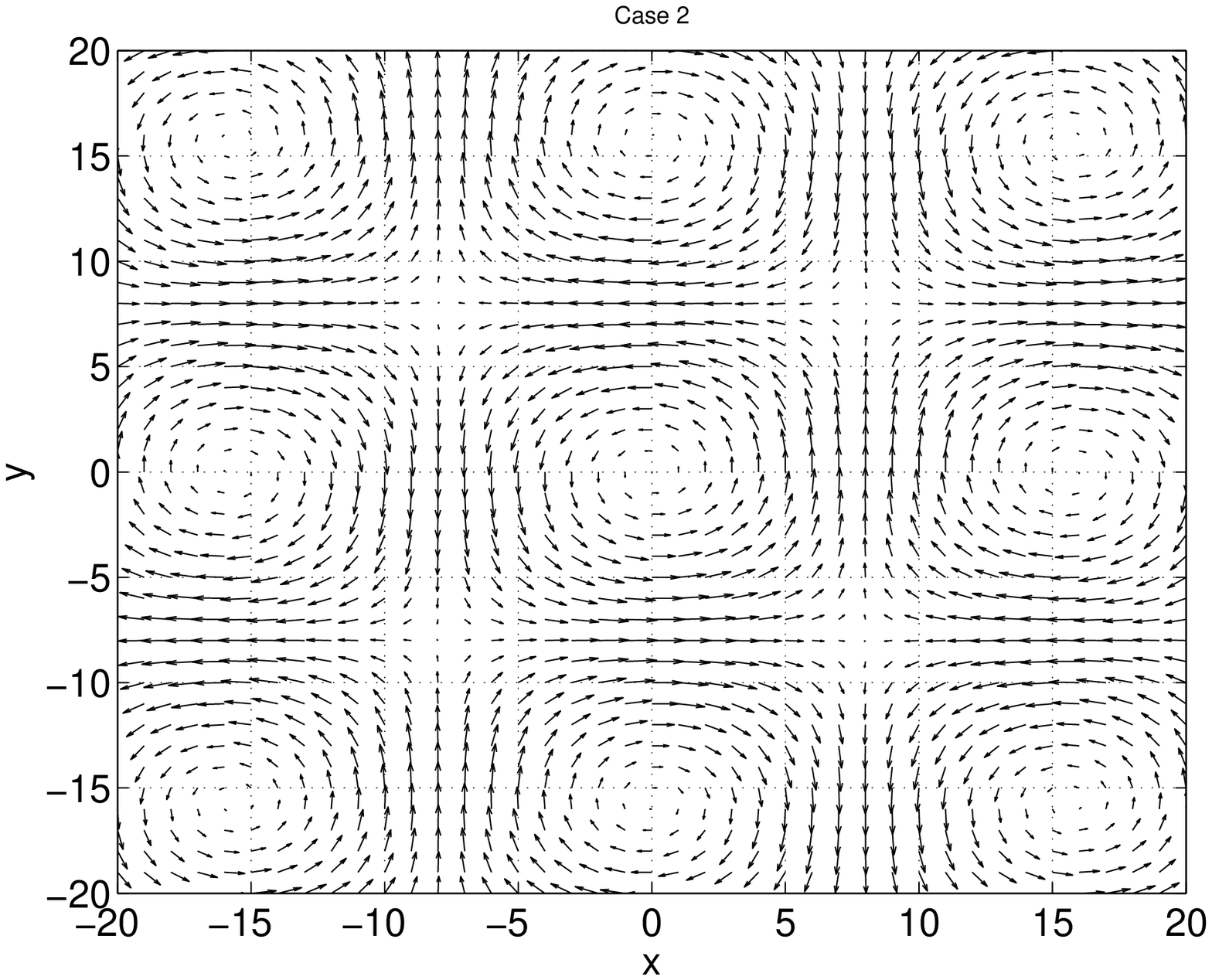}
 \end{center}
Figure 3.  Snapshots of velocity fields for Model 1 (left) and Model 2  (right)
 \end{figure}

Let $S_1$ and $S_2$ be the first explosion moments for $p(t)$ and $q(t)$ respectively, then
$$
S=\min(S_1, S_2)
$$
Certainly the knowledge of $E{S_i}$ is not enough to find $E(S)$, but the latter can be expressed in terms of
$\bar F(t)=P\left( S_1>t\right)$ as
$$
ES=\int_0^\infty \bar F(t)^2dt
$$
Manipulating with this formula it is not difficult to  present an example where the expected value of the minimum
of two independent identically distributed random variables $X,Y$ is finite while  the expectation of  each variable is infinite due to heavy tails of $X, Y$ distribution. So, theoretically speaking
$E{S_i}$ and $E(S)$ may differ in an order.  Fortunately it is not the case here since the tail of distribution of $S_1$ is exponential and can be evaluated [5].

\begin{figure}[hbp]
\begin{center}
\includegraphics[width=80mm, height=60mm]{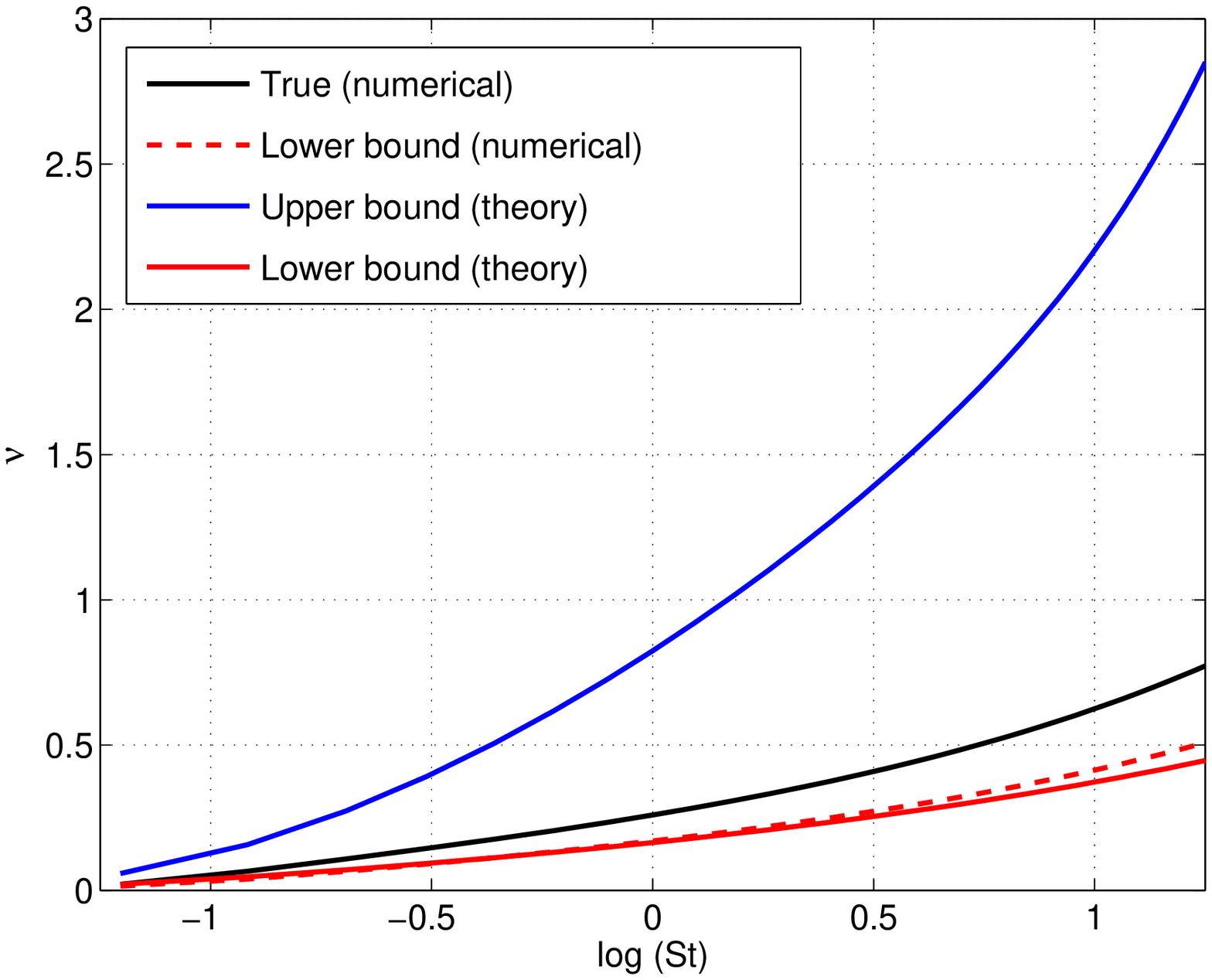}
\end{center}
Figure 4. The theoretical upper and lower bounds for the caustic rate $\nu(s)$ as functions of Stokes number $s$ and its numerical version
 \end{figure}

Namely from Theorem 1.1 in [7] it follows that the exponential moment
$$
E\left( e^{\gamma S_1}\right) <\infty  \eqno (23)
$$
is finite for all $\gamma = \gamma (s)$ satisfying
$$
\gamma(s)<\frac{1}{g(s)}  \eqno (24)
$$
and
$$
g(s)=\max\left\{x: \left(\int_{-\infty }^{x}\exp (y^{3}/3s^2+y^{2}/2s^2 )dy\right)\left(\int_{x}^{\infty }\exp (-y^{3}/3s^2-y^{2}/2s^2 )dy\right)\right\}  
$$  
Hence it is reasonable to assume that
$$
\bar F(t) \geq e^{-\gamma t}  \eqno (25)
$$
with $\gamma$ satisfying (24).
In view of this assumption
$$
E(S)> \frac {1}{2g(s)}
$$
On the other side obviously that
$$
E(S)<E(S_1)
$$
Thus for the mean number of caustics $\nu(s)= 1/E(S)$ one gets from (9)
$$
\displaystyle \left(2\pi^2 \sqrt{z}M(z)^2\right)^{-1}<\nu(s)<2g(s)
$$
For small $s$ asymptotics of the lower bound  and upper bound coincide and we get
$$
\nu(s)\approx \displaystyle \frac{1}{4\pi s^{3/2}}e^{-\frac{1}{6s^2}}
$$
while for large $s$ the corresponding asymptotics differ just by a constant
$$
0.4 s^{2/3}<\nu(s)<1.25 s^{2/3}
$$
Approximation (25) is quite speculative and indeed it greatly overestimates $\nu(s)$.  It can be seen by comparing the bounds with exact curve $\nu(s)$ obtained from solving the corresponding PDE for $\bar F(t)$ (Section 1).  The upper and lower bounds for $\nu(s)$ are shown in Figure 4  as well as  its numerical version.

Finally, notice that LE in Model 1 is given by the same expression (11) as in $1D$ case. Indeed, in this case each component of the separation process
$\Delta \mathbf r = \mathbf r(t,\mathbf a)-\mathbf r(t,\mathbf 0)$ is the separation process for one dimensional flow (6). Hence $|\Delta \mathbf r| \approx 
|\mathbf a\exp{(\lambda t)}$ for small $|\mathbf a|$ and large $t$, where $\lambda$ is LE for (6). Then our claim follows from definition (3).

In Model 2 LE cannot be found by such simple tools because the equations for $x$ and $y$ components are not split.

\bigskip

{\large \textbf{5. Isotropic forcing}}

\bigskip

Assume that in the original equations the forcing is isotropic, then its covariances are given by [2]
\begin{eqnarray*}
b^{11}(x,y) &=&b_{N}(r)+\frac{x^{2}}{r^{2}}(b_{L}(r)-b_{N}(r)),\quad
b^{22}(x,y)=b_{N}(r)+\frac{y^{2}}{r^{2}}(b_{L}(r)-b_{N}(r)), \\
b^{12}(x,y) &=&b^{21}(x,y)=\frac{xy}{r^{2}}(b_{L}(r)-b_{N}(r)),
\end{eqnarray*}
where $r=\sqrt{x^{2}+y^{2}}$. Assume smooth longitudinal and normal correlation functions 
\begin{eqnarray*}
b_{L}(r) &=&b_{0}-\frac{1}{2}\beta _{L}r^{2}+O(r^{4}), \\
b_{N}(r) &=&b_{0}-\frac{1}{2}\beta _{N}r^{2}+O(r^{4}),
\end{eqnarray*}
with $b_{0},\beta _{L},\beta _{N}>0$. Introduce $s_L=\sqrt{2\beta_L\tau^3},\ s_N=\sqrt{2\beta_N\tau^3}$

Then from (17) it can be derived that (15) takes form
\begin{equation*}
\begin{array}{lll}
dm_{1} & = & (-m_{1}+(m_{4}^{2}-m_{1}^{2}-m_{2}^{2}-m_{3}^{2})/2)dt+s_Ldw_1, \\
dm_{2} & = & (-m_{1}m_{2}-m_{2})dt+s_Ndw_2, \\
dm_{3} & = & (-m_{1}m_{3}-m_{3})dt + s_Ldw_3, \\
dm_{4} & = & (-m_{1}m_{4}-m_{4})dt+s_Ndw_4,\quad
\end{array}
\end{equation*}
here
$w_j,\ \ j=1,2,3,4$ are independent standard Wiener processes.

No essential progress is made  in analyzing this system yet because none of variables can be eliminated. The only reason for presenting it here is a great importance of the isotropic case for applications.

\bigskip

{\large \textbf{6. Conclusion and Discussion}}

\bigskip
While one dimensional stochastic models for inertial particles have been comprehensively studied, [1,8] there was a lack of $2D$ examples with a significant analytical advances. Here we suggested two such examples, where the mean time  to explosion in the particle density , $\mu (s)$, can be analytically estimated for the full range of Stokes number $s$ and can be accurately evaluated by solving an initial/boundary problem for a one dimensional parabolic equation.
The reported advance in the proposed models is due to reduction in the number of unknowns in the system (15) from four to two. Alas, for the most interesting isotropic case such a reduction is not possible, but a hope for finding asymptotics of $\nu(s)$ still remain.
For both models the reciprocal $\nu(s)=1/\mu(s)$, interpreted as the intensity of caustics,  increases monotonically from zero to infinity that is not a surprise at all.

Regarding to a Riemann type equation for the underlying velocity field (4), $\mu(s)$ can be viewed as a time scale on which a solution remains unique. It should be recognized that the interpretation of $\mu$ and $\nu$ in terms of the Eulerian velocity field is not clear enough except the deterministic case with the initial velocity field $\mathbf u_0(\mathbf x)$ for which $|\partial \mathbf u_0 \partial \mathbf x/|$ and $\nabla \cdot \mathbf u_0$ do not depend on $\mathbf x$ at all.

\bigskip

{\large \textbf{Appendix}}

\smallskip

{ \textbf{A1. Proof of Proposition 2}}

\smallskip

The proof is based on the following statement

\textbf{Lemma }. \textit{If there exist real} $m$ \textit{and smooth
bounded function} $V(x)$ \textit{such that}
$$
\quad V(\infty )=V(-\infty )=0  \eqno (A1)
$$
\textit{and}
$$
\quad LV(x)=x-m, \eqno (A2)
$$
\textit{\ then with probability one}

$$
\bar p=m 
$$

\textbf{Proof }. From (A2) and Ito formula it follows that $V(p(t))$ satisfies 
\[
\quad dV=(p-m)dt+sV^{\prime }dw.
\]
By integrating both sides we get
$$
\frac{1}{T}(V(p(T))-V(p(0))=\frac 1T \int_0^Tp(t)dt-m+\frac{s}{T}\int_{0}^{T}V^{\prime }(p(t))dw \eqno (A3)
$$
The second term on the right hand side goes to zero as $T\to\infty$ due to  the large numbers law. The difference on the left hand side (LHS) of (A4) also converges to zero because of the boundness of $V$ and condition (A1).
Lemma is proven.

Notice importance of (A1). If  $\quad V(\infty )\neq V(-\infty )$ then the jumps of $V$ at explosion moments can
accumulate leading to a non-zero limit of LHS.

Now we directly construct such a function $V(x)$ and determine $m$.

First, we take a bounded solution of (A2) 
$$
V(x)=\displaystyle \frac{1}{s^2}\int_{-\infty }^{x}\exp \left(\frac{y^{3}/3+y^{2}/2 }{s^2}%
\right)\int_{y}^{\infty }\exp \left(-\frac{u^{3}/3+u^{2}/2 }{s^2}\right)(u-m)dudy.
$$
for which $V(-\infty)=0$. To ensure (A1) one should set
$$
m=\frac MN
$$
where
$$
\begin{array}{lll}
M=\int_{-\infty }^{\infty}\exp \left(\frac{y^{3}/3+y^{2}/2 }{s^2}%
\right)\int_{y}^{\infty }\exp \left(-\frac{u^{3}/3+u^{2}/2 }{s^2}\right)ududy\\ \\
N=\int_{-\infty }^{\infty}\exp \left(\frac{y^{3}/3+y^{2}/2 }{s^2}%
\right)\int_{y}^{\infty }\exp \left(-\frac{u^{3}/3+u^{2}/2 }{s^2}\right)dudy 
\end{array}\eqno (A4)
$$
The integral for $M$ is meant as Cauchy principal value. That allows for changing the order of integration after substitution
\[
t=u-y,\ \ y'=ys^{2/3}
\]
Integration in $y'$ leads to
\[
\quad m=\frac 12\left( -1+\frac{g^{\prime }(z)}{2\sqrt{z}g(z)}\right)
\]
where
\[
\quad g(z)=\int_{0}^{\infty }\exp (zt-t^{3}/12)\frac{dt}{\sqrt{t}}.
\]
To complete the proof 
 one should account for, [5],
\[
\quad g(x)=\pi ^{-1}M(x)^{2}.
\]
This relation also can be derived from the fact that the both sides satisfy
the same differential equation and same initial conditions, [9].

Finally, notice that the expression for $m$ after changing the order of integration in (A4) turns to  the mean of $p(t)$ with respect to the invariant measure $\pi (p)$ given in (12)

Proposition 2 is proven.  Proposition 3 can be proven by the same arguments.

\bigskip

{ \textbf{A2. Details of the computational algorithm for solving (10)}}

\smallskip
Let $h$ and $\Delta t$ be space and time respectively, $X>0$ a big enough number, $x_i=-X+(i-1)h,\ \ n=(n-1)\Delta t$ the corresponding space/time grid, $i=1,...,M,\ \ n=1,...,N$ where $h=2X/M, \ \ \Delta t= t/N$. Set $b_i=\Delta t(x_i+x_i^2+s^2/h),\ \ a=\Delta t s^2/h$. Then a standard Eulerian scheme for (23) is written as
$$
\begin{array}{lll}
u_{i+1}(n)=u_i(n)+hy_i(n)\\ \\
u_{i+1}(n+1)-u_{i+1}(n) =ay_{i+1}(n)-b_iy_{i}(n)
\end {array}\eqno (A5)
$$
Notice that (A5) is well posed because $u_i(1)=1, \ i=1,...,M,\ \ \ u_1(n)=0,  \ \ y_M(n)=0,\ \ n=1,...,N$

The choice of parameters $X, h, \Delta t$ was dictated by standard conditions on ratio $\Delta t/h$ and constraints
$$
\sum_1^N \bar F(n)\Delta t \approx \mu (s)
$$
where $\mu (s) =E(S)$ is the exact value of the expectation of $S$ found from (9). In particular $\mu(1/\sqrt{2}$ =8.7735 while under our choice of $X=7.5, \Delta t= 0.001, M=61$ it was obtained that 
$$
\sum_1^N \bar F(n)\Delta t \approx 8.7785
$$
A similar comparison for other $s$ can be seen from Figure 4

\bigskip

{ \textbf{A3. Derivation of (15)}}

\smallskip

For the purpose of nondimensionalizing let us redenote the forcing on RHS of (14) by capital letters $W^{(i)}(t,x,y),\ \ i=1,2$. 
Then by differentiating (14) in $a$ and $b$ we get
\begin{eqnarray*}
du_a=-(u_a/\tau )dt+x_ad\xi_1+y_ad\xi_2, \quad dv_a=-(v_a/\tau )dt+x_ad\xi_1+y_ad\xi_4,\quad dx_a=u_adt,\quad dy_a=v_adt\\ \\
du_b=-(u_b/\tau )dt+x_bd\xi_1+y_bd\xi_2, \quad dv_b=-(v_b/\tau )dt+x_bd\xi_1+y_bd\xi_4,\quad dx_b=u_adt,\quad dy_b=v_bdt
\end{eqnarray*}
where
$$
\xi_1 =\partial W^{(1)}/\partial x,\quad \xi_2 =\partial W^{(1)}/\partial y,\quad \xi_3 =\partial W^{(2)}/\partial x,\quad \xi_4 =\partial W^{(2)}/\partial y
$$
and the covariance matrix of $\xi$'s is given by
\begin{equation*}
\left(
\begin{tabular}{llll}
$-B^{11}_{xx}$ & $0 $& $-B^{12}_{xx}$ & $-B^{12}_{xy}$ \\ \\
$0$ &  $-B^{11}_{yy}$ & $-B^{12}_{xy}$ & $-B^{12}_{yy}$\\ \\
$-B^{12}_{xx}$ & $ -B^{12}_{xy}$ & $-B^{22}_{xx}$ & $0$ \\ \\
$-B^{12}_{xy}$ & $-B^{12}_{yy}$ & $0$ & $-B^{22}_{yy}$ \\ \\
\end{tabular}
\right)
\end{equation*}
where $B^{ij}$ is the space covariance matrix of $\left(W^{(1)}, W^{(2)}\right)$

Introduce
\begin{eqnarray*}
J_{1}(t) &\equiv &\frac{\partial x}{\partial a}\frac{\partial v}{\partial b}-%
\frac{\partial x}{\partial b}\frac{\partial v}{\partial a}, \\
J_{2}(t) &\equiv &\frac{\partial y}{\partial a}\frac{\partial u}{\partial b}-%
\frac{\partial y}{\partial b}\frac{\partial u}{\partial a}, \\
J_{3}(t) &\equiv &\frac{\partial x}{\partial a}\frac{\partial u}{\partial b}-%
\frac{\partial x}{\partial b}\frac{\partial u}{\partial a}, \\
J_{4}(t) &\equiv &\frac{\partial y}{\partial a}\frac{\partial v}{\partial b}-%
\frac{\partial y}{\partial b}\frac{\partial v}{\partial a}, \\
J_{5}(t) &\equiv &\frac{\partial u}{\partial a}\frac{\partial v}{\partial b}-%
\frac{\partial u}{\partial b}\frac{\partial v}{\partial a},
\end{eqnarray*}
Applying Ito formula we obtain
\begin{eqnarray*}
dJ &=&(J_{1}-J_{2})dt,\quad ~ \\
dJ_{1} &=&(-J_{1}/\tau +J_{5})dt+Jd\xi_4, \\
dJ_{2} &=&(-J_{2}/\tau -J{5})dt+Jd\xi_1, \\
dJ_{3} &=&(-J_{3}/\tau )dt+Jd\xi_2, \\
dJ_{4} &=&(-J_{4}/\tau )dt+Jd\xi_3, \\
dJ_{5} &=&(-J_{5}/\tau )dt+J_1d\xi_1+J_2d\xi_2-J_3d\xi_3-J_4d\xi_4,
\end{eqnarray*}
To proceed to dimensionless variables assume 
$$
W^{(i)}(t,x,y)=\sigma\tau^{1/2} w^{(i)}(t/\tau, x/L,y/L),\ \ i=1,2
$$
where $ w^{(i)}$ are dimensionless Wiener processes, thereby in dimensionless variables
$$
Edw^{(i)}(t,0,0)dw^{(j)}(t,x,y)= b^{ij}(x,y)dt
$$
with
$$
B^{ij}(x,y)=\sigma^2b^{ij}(x/L,y/L)
$$
Introduce a Stokes number similarly to (8)
$$
s=\sigma\tau^{3/2}/L
$$
The result is
\begin{equation*}
\begin{array}{lll}
dJ & = & I_{1}dt,\quad ~ \\
dI_{1} & = & (-I_{1} +2J_{5})dt+sJd\eta_1, \\
dI_{2} & = & -I_{2} dt+sJd\eta_2, \\
dI_{3} & = & -I_{3}dt+sJd\eta_3, \\
dI_{4} & = & -I_{4}dt+sJd\eta_4
\end{array}
\end{equation*}
\begin{equation*}
 I_{1}=\tau( J_{1}-J_{2}),\quad I_{2}=
\tau(J_{1}+J_{2}),\quad I_{3}=\tau( J_{4}-J_{3}),\quad I_{4}=
\tau(J_{3}+J_{4})
\end{equation*}
where
$$
\eta_1=(\xi_4 -\xi_1)/h, \eta_2=(\xi_4 +\xi_1)/h, \eta_3=(\xi_3 -\xi_2)/h,\eta_4= (\xi_3 +\xi_2)/h,\ \ \ h=\sigma\tau^{1/2}
$$
We do not need a stochastic equation for $J_5$ since it can be found from the following equation
\begin{equation*}
I_{2}^{2}-I_{1}^{2}-I_{4}^{2}+I_{3}^{2}=-4JJ_{5},\eqno(A6)
\end{equation*}
which follows from
\begin{equation*}
J_{1}J_{2}+JJ_{5}=J_{3}J_{4}.
\end{equation*}

We can reduce the number of freedom degrees by introducing
 $m_{k}=I_{k}/J,\quad k=1,..., 5$
where $m_{5}$ is eliminated by means of (A6). The result is system (15)

\bigskip

\textbf{\Large References}

\bigskip

\begin{description}

\item  1. K. Gustavsson and B. Mehlig, Statistical models for spatial patterns of inertial particles in turbulence, 
\emph{arXiv:1412.4374v1 [physics.flu-dyn]},  2014.

\item  2. {A. S. Monin and A.M. Yaglom, ''Statistical Fluid Mechanics:
Mechanics of Turbulence'', \ MIT Press, Cambridge, MA, 1975. }

\item  3. {L. I. Piterbarg,  The top Lyapunov exponent for a stochastic flow
modeling the upper ocean turbulence. \emph{SIAM J. Appl. Math.}, \textbf{62}%
, 777-800, 2001.}

\item  4. I. Karatzas and S.E. Shreve, \emph{Brownian Motion and Stochastic
Calculus}, Springer-Verlag., NY-Berlin-Heidelberg, 1991.

\item  5. B. I.  Halperin, Green's functions for a particle in a
one-dimensional random potential, \emph{Phys. Rev.,} 139, A104-117, 1965.

\item  6. I. Karatzas I. and S.E. Shreve, Distribution of the time to explosion for one-dimensional diffusions 
\emph{Probab. Theory Relat. Fields}, 164, 1027-1069, 2016.

\item  7. O. Loukianov, D. Loukianova, and S. Song, Spectral gaps and exponential integrability of hitting times for linear diffusions, 
\emph{Ann. Inst. H. Poincaré Probab. Statist.
Volume 47, Number 3 , 679-698., }, 2011.

\item 8. G. Falkovich , S. Musacchio , L. Piterbarg , and M. Vucelja  (2007) Inertial particles driven by a telegraph noise, \emph{ Physical Review E }, 76, 2007

\item 9.   M. Abramowitz and I. A. Stegun, \emph{ Handbook of Mathematical Functions} Dover Publications, INC., New York, 1970

\end{description}

\end{document}